\documentclass[twocolumn,aps,prd,10pt]{revtex4-2}
\usepackage{bm}
\usepackage{amsfonts}
\usepackage{amssymb}
\usepackage{amsmath}
\usepackage{graphicx}
\begin{document}

\title{Helical beams of electrons in a magnetic field:\\
New analytic solutions of the Schr\"odinger and Dirac equations}
\author{Iwo Bialynicki-Birula}\email{birula@cft.edu.pl}
\affiliation{Center for Theoretical Physics, Polish Academy of Sciences\\
Aleja Lotnik\'ow 32/46, 02-668 Warsaw, Poland}
\author{Zofia Bialynicka-Birula}
\affiliation{Institute of Physics, Polish Academy of Sciences\\
Aleja Lotnik\'ow 32/46, 02-668 Warsaw, Poland}
\date{\today}

\begin{abstract}
We derive new solutions of the Schr\"odinger, Klein-Gordon and Dirac equations which describe the motion of  particles in a uniform magnetic field. In contrast to the well known stationary solutions, our solutions exhibit the behavior of quantum particles which very closely resembles classical helical trajectories. These solutions also serve as an illustration of the meaning of the Ehrenfest theorem in relativistic quantum mechanics.
\end{abstract}

\maketitle

\section{Introduction}

According to classical mechanics charged particles move in a uniform magnetic field along helical trajectories (Fig.~1). In this work we describe new solutions of the Schr\"odinger, the Klein-Gordon and the Dirac equations which are close counterparts of the classical trajectories.

\begin{figure}[b]
\begin{center}
\vspace{0.35cm}
\includegraphics[width=0.3\textwidth,
height=0.25\textheight]{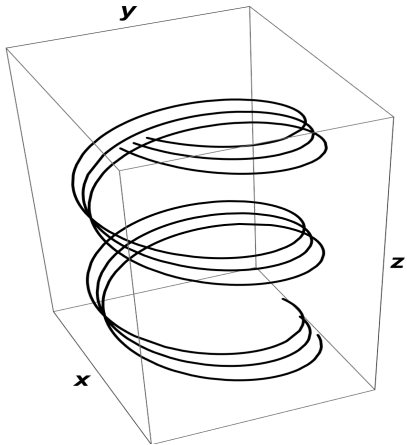}
\caption{Helical trajectories of charged particles in a uniform magnetic field for three choices of initial conditions.} \label{fig1}
\end{center}
\end{figure}

Analytic solutions of wave equations in quantum mechanics are most often obtained by the separation of variables and the time variable is usually the first variable which is subjected to this procedure. In this way we obtain the stationary solutions -- energy eigenstates. The wave functions obtained in this way do not resemble their classical counterparts because the probability density does not depend on time -- nothing moves. For example, the wave functions of stationary states in the Coulomb potential and the elliptical Keplerian orbits do not seem to have anything in common. In principle, it should be possible to exhibit the classical-quantum relation by superposing Coulombic wave functions with different orientations of the $z$ axis and with different phases but this very cumbersome task has never been accomplished.

Analytic solutions of the Schr\"odinger and Dirac equations for a charged particle in a uniform magnetic field have been known for almost hundred years \cite{rabi,fock}. They have been described in detail in \cite{bg}. However, these are all {\em stationary solutions} so that the probability density does not move. Therefore, they cannot be viewed as quantum counterparts of the classical helical trajectories depicted in Fig.1.

Our method of constructing new solutions of the Schr\"odinger and Dirac equations employs the procedure that might be called the injection of classical trajectories (ICT) into the wave functions. This method has been applied to the solutions of the Gross-Pitaevskii equation \cite{gpv} and to the center of mass motion of Bose-Einstein condensate in a harmonic trap \cite{bb}. In the present work we apply this method to the motion of electrons in a magnetic field both in the nonrelativistic and in the relativistic domain. The states of quantum particles described by the wave functions obtained here are as close as possible to their classical counterparts; the centers of the wave packets follow the classical trajectories.

\section{Classical dynamics}

The classical trajectories will be described here within the canonical Hamiltonian formalism because the classical trajectories used in the ICT method are most easily expressed in terms of canonical variables,
\begin{align}\label{cv}
\bm{p}=\{p_x,p_y,p_z\},\quad \bm{r}=\{x,y,z\}.
\end{align}
Moreover, in this formalism the equations of motion have the same form in the nonrelativistic and in the relativistic case.

The starting point are the formulas for the Hamiltonian describing the electron in a constant magnetic induction field $\bm B$ in the nonrelativistic and in the relativistic case,
\begin{align}
H_{\rm NR}&=\frac{1}{2m}\left(\bm{p}-e{\bm A}\right)^2,\label{ham1}\\
H_{\rm RL}&=c\sqrt{m^2 c^2+\left(\bm{p}-e{\bm A}\right)^2}.\label{ham2}
\end{align}
We found it convenient to use the geometrical units $\rm{1/meter^2}$ for the magnetic induction. This means that in our notation $B$ is related to its counterpart in the SI units by the formula $B=eB_{\rm SI}/\hbar$.

In the case of a constant vector $\bm B$ in the symmetric gauge we have,
\begin{align}
e{\bm A}=\frac{\hbar}{2}{\bm B}\times{\bm r}.
\end{align}
If the $z$ axis is chosen in the direction of $\bm B$, the Hamiltonians (\ref{ham1}) and (\ref{ham2}) take on the form,
\begin{align}
\!\!\!H_{\rm NR}=\!\frac{1}{2m}\!&\left[\bm{p}^2+\frac{(\hbar B)^2}{4}(x^2\!+y^2)-\!\hbar B(x p_y-y p_x)\right],\label{ham1a}\\
&H_{\rm RL}=c\sqrt{m^2 c^2+2mH_{\rm NR}}.\label{ham2a}
\end{align}

The canonical equations of motion in the nonrelativistic and in the relativistic case have the same form,
\begin{subequations}\label{ceq}
\begin{align}
\frac{dx(t)}{dt}&=\frac{p_x(t)}{M}+\frac{\hbar\,B}{2M}y(t),\\
\frac{dy(t)}{dt}&=\frac{p_y(t)}{M}-\frac{\hbar\,B}{2M}x(t),\\
\frac{dz(t)}{dt}&=\frac{p_z(t)}{M},\\
\frac{dp_x(t)}{dt}&=-\frac{(\hbar B)^2x(t)}{4M}+\frac{\hbar\,B}{2M}p_y(t),\\
\frac{dp_y(t)}{dt}&=-\frac{(\hbar B)^2y(t)}{4M}-\frac{\hbar\,B}{2M}p_x(t),\\
\frac{dp_z(t)}{dt}&=0.
\end{align}
\end{subequations}
The only difference is in the meaning of the parameter $M$. In the nonrelativistic case, $M$ is simply the rest mass of the particle ($M=m$), while in the relativistic case $M$ denotes the relativistic mass ($M=H_{\rm RL}/c^2$), which is a constant of motion, although its value depends on the initial conditions. The solutions of Eqs.~(\ref{ceq}) are,
\begin{widetext}
\begin{subequations}\label{sol}
\begin{align}
x(t)&=\frac{1}{2}x_0(1+\cos(\omega t))+\frac{1}{2}y_0\sin(\omega t)+\frac{p_{x0}\sin(\omega t)}{M\omega}+\frac{p_{y0}(1-\cos(\omega t))}{M\omega},\\
y(t)&=-\frac{1}{2}x_0\sin(\omega t)+\frac{1}{2}y_0(1+\cos(\omega t))-\frac{p_{x0}(1-\cos(\omega t))}{M\omega}
+\frac{p_{y0}\sin(\omega t)}{M\omega},\\
z(t)&=z_0+\frac{p_{z0}}{M}t,\\
p_x(t)&=-\frac{1}{4}x_0M\omega\sin(\omega t)-\frac{1}{4}y_0M\omega(1-\cos(\omega t))+\frac{1}{2}p_{x0}(1+\cos(\omega t))+\frac{1}{2}p_{y0}\sin(\omega t),\\
p_y(t)&=\frac{1}{4}x_0M\omega(1-\cos(\omega t))-\frac{1}{4} y_0m\omega\sin(\omega t)-\frac{1}{2}p_{x0}\sin(\omega t)+\frac{1}{2}p_{y0}(1+\cos(\omega t)),\\
p_z(t)&=p_{z0},
\end{align}
\end{subequations}
\end{widetext}
where $\omega=\hbar B/M$ is the cyclotron frequency and the parameters with the subscript 0 denote the initial values of positions and momenta. The trajectories in Fig.1 depict these solutions.

\section{Helical solutions of the Schr\"odinger equation}

We begin with the construction of helical beams of the electrons described by the solutions of the Schr\"odinger equation. These solutions will also be used in the next Section in the relativistic theory. In order to treat the motion in a magnetic field, we need an extension of the ICT method used in \cite{gpv,bb} to cover the case of the Hamiltonian (\ref{ham1a}) which in addition to the harmonic potential contains also the part with positions and momenta. The extended version of the ICT method can be stated as the following theorem.

Let us consider the Schr\"odinger equation with a general quadratic Hamiltonian operator,
\begin{align}\label{qham}
\hat{H}=\frac{1}{2}\hat{p}_iA^{ij}\hat{p}_j+\frac{1}{2}\hat{x}^iB_{ij}\hat{x}^j
+\hat{p}_iC^i_{\;j}\hat{x}^j.
\end{align}
Such a Hamiltonian is obtained from its classical counterpart (\ref{ham1}) by replacing the canonical variables by the operators $\hat{x}^i$ and $\hat{p}_i$.

The theorem states that from any solution $\psi(x,y,z,t)$ of the Schr\"odinger equation with the Hamiltonian (\ref{qham}) one can obtain a family of new solutions $\psi_{\rm ICT}(x,y,z,t)$ defined by the following unitary mapping:
\begin{widetext}
\begin{align}\label{unit}
\psi_{\rm ICT}(x,y,z,t)
=\hat{U}\psi(x,y,z,t)=\exp\left[-\frac{i}{2\hbar}p_i(t)x^i(t)\right]
\exp\left[\frac{i}{\hbar}p_i(t)\hat{x}^i\right]
\exp\left[-\frac{i}{\hbar}\hat{p}_ix^i(t)\right]\psi(x,y,z,t),
\end{align}
where ($x^i(t),p_j(t)$) obey the classical equations of motion,
\end{widetext}
\begin{align}\label{eqm}
\frac{d{x}^i(t)}{dt}&=A^{ij}p_j(t)+C^i_{\,j}x^j(t),\\
\frac{d{p}_i(t)}{dt}&=-B_{ij}x^j(t)-C^j_{\,i}p_j(t).
\end{align}

The proof of the theorem (see the Appendix) consists in showing that the unitary operator $\hat{U}$ leaves the Schr\"odinger equation intact,
\begin{align}\label{proof}
\hat{U}^\dagger\left(\hat{H}-i\hbar\partial_t\right)\hat{U}
=\left(\hat{H}-i\hbar\partial_t\right).
\end{align}

The unitary operator $\hat{U}$ has three  parts: the first two are just the phase factors. The third one, which plays an essential role in our construction of helical solutions, can also be written as the displacement operator,
\begin{align}\label{dis}
\exp\left[-\frac{i}{\hbar}\hat{p}_ix^i(t)\right]
=\exp\left[-x^i(t)\partial_i\right].
\end{align}
When acting on a wave function, it displaces the position variables $x^i$ by $x^i(t)$. To put it more vividly, it injects the classical trajectory into the wave function.
\begin{align}\label{dis1}
\exp\left[-x^i(t)\partial_i\right]\psi(x^i)=\psi(x^i-x^i(t)).
\end{align}
Note that consecutive multiple injections of classical trajectories produces the same result as a single injection of the superposition of all trajectories.

We will apply now our theorem, to generate families of helical solutions of the Schr\"odinger equation,
\begin{align}\label{sch}
i\hbar\partial_t\Phi=-\frac{\hbar^2}{2m}\left[\bm{\nabla}-
\frac{ie}{\hbar}\bm{A}\right]^2\Phi,
\end{align}
from well known (cf., for example \cite{bal,bck}) stationary solutions $\psi_{nlp_z}$  for a charged particle in the uniform magnetic field,
\begin{widetext}
\begin{align}\label{lag}
\psi_{nlp_z}(x,y,z,t)=\sqrt{\frac{B^{l+1}n!}{2\pi2^l(n+l)!}}\,
\exp\left[-\frac{i(p_z^2+(2n+1)B\hbar^2)t}{2m\hbar}
+\frac{ip_z z}{\hbar}-\frac{B(x^2+y^2)}{4}\right]
\,(x+iy)^l\,L_n^l\left[\frac{B(x^2+y^2)}{2}\right],
\end{align}
\end{widetext}
\begin{figure}
\begin{center}
%\vspace{0.7cm}
\includegraphics[width=0.9\textwidth,
height=0.2\textheight]{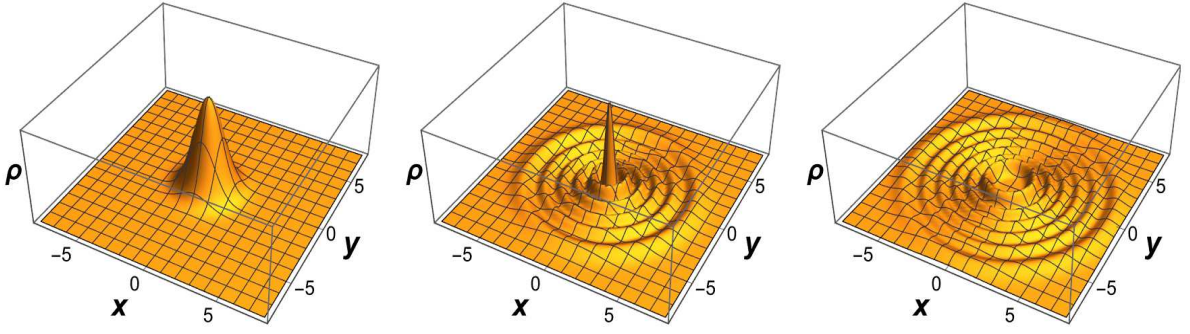}
\caption{Probability density $\rho=|\psi_{nlp_z}|^2$ as a function of $x$ and $y$ for the following choices of quantum numbers $(n,l)$: (0,0), (5,0) and (5,5). } \label{fig2}
\end{center}
\end{figure}
where $L_n^l$ are associated Laguerre polynomials \cite{math}. The probability density obtained from these solutions depicted in Fig.~2 does not show any resemblance to the helical trajectories of classical charged particles in Fig.~1. However, the functions $\psi_{nlp_z}$ can be used as building blocks in the construction of genuine helical solutions. First, note that these solutions depend on $z$ only through the phase factor $\exp(ip_zz/\hbar)$. Therefore, the application of the ICT method to these wave functions would not produce helical beams because the probability density will not depend on $z$. However, from (\ref{lag}) we can construct square-integrable wave packets $\psi_{P}$ by the integration over $p_z$ with the Gaussian weight $\exp(-d^2p_z^2/(2\hbar^2))$. The resulting wave functions, fully normalized in all three variables and localized on the $z$ axis, have the form,
\begin{widetext}
\begin{align}\label{wp}
&\psi_{P}(x,y,z,t)=\pi^{-1/4}\exp\left[-\frac{mz^2}
{2md^2+2i\hbar t}\right]
\left(d+\frac{i\hbar t}{md}\right)^{-1/2}\nonumber\\
\times\sqrt{\frac{B^{l+1}n!}{2\pi2^l(n+l)!}}\,
&\exp\left[-\frac{i(2n+1)B\hbar\,t}{2m}\right]\exp\left[-\frac{B(x^2+y^2)}{4}
\right]\,(x+iy)^l\,L_n^l
\left[\frac{B(x^2+y^2)}{2}\right].
\end{align}
\end{widetext}
These solutions of the Schr\"odinger equation will be now used to generate helical solutions.

The application of the ICT method (\ref{unit}) to the wave-packet solutions (\ref{wp}) of the Schr\"odinger equation produces the following wave functions that describe helical beams,
\begin{widetext}
\begin{align}\label{sh}
&\psi_{H}(x,y,z,t)=e^{-i\varphi}\sqrt{\frac{B^{l+1}n!}{2\pi^{3/2}2^l(n+l)!}}
\exp\left[-\frac{mz_S^2}
{2md^2+2i\hbar t}\right]
\left(d+\frac{i\hbar t}{md}\right)^{-1/2}\nonumber\\
\times
&\exp\left[-\frac{B(x_S^2+y_S^2)}{4}
\right]\,(x_S+iy_S)^l\,L_n^l
\left[\frac{B(x_S^2+y_S^2)}{2}\right],
\end{align}
where we introduced a shortened notation $x_S=x-x(t)$, etc. for shifted variables. We will not need the explicit form of the phase $\varphi$ because in what follows we will only discuss the probability density $\rho_H$,
\begin{align}\label{rho}
\rho_H(x_S,y_S,z_S,t)=&|\psi_{H}(x,y,z,t)|^2=N\exp\left[-\frac{m^2d^2z_S^2}
{m^2d^4+\hbar^2t^2}\right]
\left(d^2+\frac{\hbar^2t^2}{m^2d^2}\right)^{-1/2}\nonumber\\
\times
&\exp\left[-\frac{B(x_S^2+y_S^2)}{2}
\right]\,(x_S^2+y_S^2)^l\,L_n^l
\left[\frac{B(x_S^2+y_S^2)}{2}\right]^2.
\end{align}
\end{widetext}
The general form of the probability density makes it possible, without doing detailed calculations, to draw the conclusion that the center of the wave packet follows exactly the classical trajectory. The average values of $x, y$ and $z$ are obtained by simple shifts of the integration variables from the integrals of the density multiplied by the corresponding coordinate. For example,
\begin{align}\label{av}
\langle x\rangle&=\int\!d^3r\,x\rho_H(x-x(t),y-y(t),z-z(t),t)\nonumber\\
&=\int\!d^3r\,(x+x(t))\rho_H(x,y,z,t)=x(t).
\end{align}
The integral of $x\rho_H$ vanishes, because after the shift, $\rho_H$ is an even function of $x$. The same argumentation gives analogous results for the remaining variables,
\begin{align}\label{av1}
\langle y\rangle=y(t),\quad \langle z\rangle=z(t).
\end{align}
Therefore there are no quantum corrections to the motion of the center of the wave packet; it follows exactly the helical classical trajectories shown in Fig.1. The name helical solutions of the Schr\"odinger equation is fully justified.

\section{Helical solutions of the Dirac equation}

The direct application of the ICT method to the  Dirac equation is not possible because the Dirac Hamiltonian is not a quadratic expression in positions and momenta. To overcome this problem, in our construction we will use the Klein-Gordon equation as an intermediate step. The connection between the Klein-Gordon equation and the Dirac equation is best seen in the Weyl (chiral) representation \cite{weyl} of the Dirac matrices. This connection has been employed in \cite{bb1,bb2} to generate new solutions of the Dirac equations in free space.

The Dirac equation in the Weyl representation is a set of two coupled equations for two-dimensional relativistic spinors $\phi$ and $\chi$,
\begin{subequations}\label{weyl}
\begin{align}
i\lambdabar\left[1/c\,\partial_t+\bm{\sigma}\!\cdot\!(\bm{\nabla}
-\frac{ie}{\hbar}\bm{A})\right]\phi=\chi,\label{weyl1}\\
i\lambdabar\left[1/c\,\partial_t-\bm{\sigma}\!\cdot\!(\bm{\nabla}
-\frac{ie}{\hbar}\bm{A})\right]\chi=\phi,\label{weyl2}
\end{align}
\end{subequations}
where $\lambdabar=\hbar/(mc)$ is the reduced Compton wave length. The electromagnetic potential is chosen in the temporal gauge. The stationary solutions of the Dirac equation in the uniform magnetic field \cite{rabi} are similar to the corresponding solutions of the Schr\"odinger equation. They have to be subjected to the same procedure as their nonrelativistic counterparts before they acquire a helical form.

The spinors $\phi$ and $\chi$ satisfying the Dirac equation have the following property. If the spinor $\phi$ satisfies the Klein-Gordon equation (with the additional term describing the direct coupling of the magnetic moment to the field),
\begin{align}\label{kgeq}
\left[\frac{1}{c^2}\partial_t^2-(\bm{\nabla}-\frac{ie}{\hbar}\bm{A})^2
+\left(\frac{mc}{\hbar}\right)^2\!\!
-\frac{e}{\hbar}\bm{\sigma}\!\cdot\!(\bm{B}-i\frac{\bm{E}}{c})\right]\phi=0,
\end{align}
then the spinor $\chi$ defined by the equation (\ref{weyl1}) automatically satisfies the equation (\ref{weyl2}). The explicit formulas for the bispinor which satisfies the Dirac equation are (spin up),
\begin{align}\label{kg2diru}
\Psi_D^{u}=\left[\begin{array}{c}\phi\\
0\\
i\lambdabar(1/c\partial_t+\partial_z-ie/\hbar A_z)\phi\\
i\lambdabar(\partial_x-ie\hbar A_x+i\partial_y+e/\hbar A_y)\phi
\end{array}\right],
\end{align}
(spin down),
\begin{align}\label{kg2dird}
\Psi_D^{d}=\left[\begin{array}{c}0\\
\phi\\
i\lambdabar(\partial_x-ie\hbar A_x-i\partial_y-e/\hbar A_y)\phi\\
i\lambdabar(1/c\partial_t-\partial_z+ie/\hbar A_z)\phi\end{array}\right].
\end{align}
These formulas have has been used by us in \cite{bb1,bb2} for the Dirac equation in free space. The construction of the solutions of the Dirac equation from the solutions of the Klein-Gordon equation works also in the presence of the electromagnetic field. In this way the problem of solving the set of two coupled equations (\ref{weyl}) is reduced to the problem of solving one equation. The use of the solutions of the Klein-Gordon equation as a stepping stone is particularly well suited in the present case, because this equation is of the second order in space derivatives and the ICT method will work.

Helical solutions of the Dirac equation will be obtained in several steps. First, we transform the Klein-Gordon equation for spin up to the form,
\begin{widetext}
\begin{align}\label{kg0}
\left[\frac{4}{c^2}\partial_+\partial_--\left(\partial_x+iBy/2\right)^2-
\left(\partial_y-iBx/2\right)^2+\left(\frac{mc}{\hbar}\right)^2- B\right]\Psi_{KG}^u=0,
\end{align}
\end{widetext}
where
\begin{align}\label{var}
t_\pm=t\pm z/c\;\;\rm{and}\;\;\partial_\pm=\partial/\partial{t_{\pm}}.
\end{align}

In order to obtain helical solutions by the ITC method, we convert the equation (\ref{kg0}) into the Schr\"odinger-like form. This can be done by restricting the functions $\Psi_D^u$ to those whose dependence on time $t_+$ is harmonic with the frequency $\omega=Mc^2/\hbar$. We also extract from $\Psi$ an additional factor harmonic in $t_-$,
\begin{align}\label{harm}
\Psi_{KG}^u=\exp\left[-\frac{i\left(M^2c^2t_++m^2c^2t_--\hbar^2 B t_-\right)}{2M\hbar}
\right]\Phi.
\end{align}
The equation satisfied by $\Phi$ has the form of the Schr\"odinger equation in two spatial dimensions with $t_-$ playing the role of time,
\begin{widetext}
\begin{align}\label{fin}
i\hbar\partial_-\Phi=-\frac{\hbar^2}{2M}\left[\partial_x^2+\partial_y^2
-\frac{B^2(x^2+y^2)}{4}-iB(x\partial_y-y\partial_x)
\right]\Phi.
\end{align}
\end{widetext}
Now we have at our disposal all stationary solutions (\ref{lag}) of the Schr\"odinger equation. The only difference is a change in notation: $p_z$ is put equal to 0, $t$ is replaced by $t_-$, and $m$ is replaced by $M$. Our ICT theorem can now be applied to generate the solutions with injected classical trajectories $x(t_-)$ and $y(t_-)$ defined by (\ref{sol}). The resulting wave function $\Psi_{KG}^{nl}$ satisfying the Klein-Gordon equation is,
\begin{align}\label{psi}
\!\!\Psi_{KG}^{nl}=e^{-i\varphi}C_n^lL_n^l\left[B/2(x_S^2+y_S^2)\right],
\end{align}
where
\begin{widetext}
\begin{align}\label{cnl}
\varphi=\frac{M^2c^2t_++\left(m^2c^2+2n\hbar^2 B \right)t_-}{2M\hbar}&+\frac{1}{2\hbar}
\left(x(t_-)p_x(t_-)+y(t_-)p_y(t_-)\right)-
\frac{1}{\hbar}\left(xp_x(t_-)+yp_y(t_-)\right),\\
C_n^l&=\exp[-i\varphi(t_-)]\exp\left[-B/4(x_S^2+y_S^2)\right]
\,(x_S+iy_S)^l.
\end{align}
\end{widetext}
We may visualize these wave functions by going back to Fig.~2 and applying the following transformations. First, we shift the shapes depicted there away from the center. Next, we start rotating them around the center with the cyclotron frequency. Finally, we move this structure with the speed of light along the $z$ axis creating the helix.

We started our construction of the function $\Psi_{KG}^{nl}$ with the solutions of the Klein-Gordon equation in the form (\ref{harm}). This construction guarantees that $\Psi_{KG}^{nl}$ contains only the positive energy parts, as required if we want to describe electrons with no parts describing positrons. On the other hand the ICT method injects classical trajectories which clearly introduce time dependence with both signs of the frequency due to the presence of trigonometric functions. It may seem that this may introduce parts describing antiparticles. We checked that in the function $\Psi_{KG}^{nl}$ all terms with the opposite sign cancel out. It must be so on physical grounds; the homogeneous magnetic field cannot supply the energy.

The requirement of positive energy is often ignored. For example, the solution of the Dirac equation in the presence of a plane electromagnetic wave (Wolkow wave function \cite{wolk}) contains terms with both signs of the frequency. Therefore, it does not describe the state of an electron but it has a significant positron component \cite{bb3}.

Our general solution of the Klein-Gordon equation $\Psi_{KG}^{nl}$ depends on 5 arbitrary parameters (in addition to two quantum numbers $n$ and $l$): on $M$ and on the initial values of the injected trajectory (\ref{sol}). Therefore, $\Psi_{KG}^{nl}$ may serve as the generating function for the plethora of various solutions because all linear operations on the wave function (for example, differentiation or integration with respect to the parameters) will produce different solutions. From all these solutions of the Klein-Gordon equation we can generate the solutions of the Dirac equation according to the formulas (\ref{kg2diru}) and  (\ref{kg2dird}).

Since the function $\Psi_{KG}^{nl}$ depends on a free parameter $M$ we could construct the wave packets analogous to those for the Schr\"odinger equation (\ref{wp}) by integrating over $M$ with some weight function. However, such wave packets are not interesting because the integration over $M$ obfuscates the helical form of the solutions. This is so because $M$ defines the cyclotron frequency and the integration over frequencies blurs the helical motion. In the nonrelativistic theory the construction of helical wave packets is possible because in this approximation the cyclotron frequency is fixed; it does not depend on the energy.

\begin{figure}
\begin{center}
%\vspace{0.7cm}
\includegraphics[width=0.2\textwidth,
height=0.2\textheight]{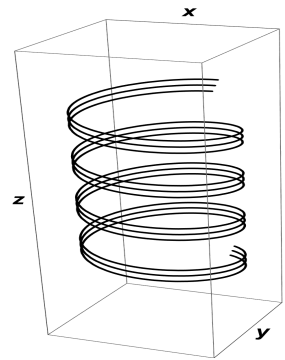}
\caption{Helical lines representing the electron beam at a given time, for three consecutive values of time. At each time the beam extends over all values of $z$. The beam is evolving in time by a helical motion.}\label{fig3}
\end{center}
\end{figure}
In the last step of the construction of the helical Dirac wave function we generate the bispinor $\Psi_D$ according to the formula (\ref{kg2diru}),
\begin{align}\label{diru}
\Psi_D^{nl}=e^{-i\varphi}C_n^l\left[\begin{array}{c}L_n^l\\
0\\
\frac{M}{m}L_n^l\\
c_1L_n^l+c_2L_{n-1}^{l+1}
\end{array}\right],
\end{align}
where
\begin{align}\label{c12}
c_1&=-\frac{1}{2mc}\left[B\hbar y(t_-)+2p_x-i(B\hbar x(t_-)-2p_y(t_-))\right],\nonumber\\
c_2&=-\frac{i}{mc}\left[B\hbar(x_S+iy_S)\right].
\end{align}
To condense the notation we omitted the argument $B(x_S^2+y_S^2)/2$ of the Laguerre polynomials. The probability density associated with the Dirac bispinor (\ref{diru}) is,
\begin{widetext}
\begin{align}\label{rhodir}
\rho_D^{nl}=\frac{N}{m^2c^2}\exp\left[-B/2(x_S^2+y_S^2)
\right](x_S^2+y_S^2)^l\left[d_1(L_n^l)^2+d_2(L_{n-1}^{l+1})^2
+d_3L_n^lL_{n-1}^{l+1}\right],
\end{align}
where $N$ is the normalization coefficient and
\begin{align}\label{d123}
d_1&=\frac{1}{m^2c^2}\left[(m^2+M^2)c^2+(B\hbar x(t_-)-2p_y(t_-))^2/4+(B\hbar y(t_-)+2p_x(t_-))^2/4\right],\nonumber\\
d_2&=\frac{B^2\hbar^2}{m^2c^2}\left[x_S^2+y_S^2\right],\qquad d_3=-\frac{B\hbar}{m^2c^2}\left[x_S(B\hbar x(t_-)-2p_y(t_-)) +y_S(B\hbar y(t_-)+2p_x(t_-))\right].
\end{align}
\end{widetext}
In all cases, except when $n=0=l$, the last term in (\ref{rhodir}) is different from zero. Since $d_3$ contains terms linear in $x_S$ and $y_S$, the average values $\langle x_S\rangle$ and $\langle y_S\rangle$ do not vanish. This produces corrections to the classical orbit. These corrections do not depend on $l$ and their dependence on $n$ is shown in the following formulas,
\begin{widetext}
\begin{subequations}
\begin{align}\label{corr}
\langle x_S\rangle=&\frac{4n\hbar(\hbar Bx(t_-)-2p_y(t_-))}{4(m^2+M^2)c^2+8n\hbar^2B
+(\hbar Bx(t_-)-2p_y(t_-))^2+(\hbar By(t_-)+2p_x(t_-))^2},\\
\langle y_S\rangle=&\frac{4n\hbar(\hbar By(t_-)+2p_x(t_-))}{(m^2+M^2)c^2+8n\hbar^2B
+(\hbar Bx(t_-)-2p_y(t_-))^2+(\hbar By(t_-)+2p_x(t_-))^2}.
\end{align}
\end{subequations}
\end{widetext}
The total average values $\langle x\rangle$ and $\langle y\rangle$ are the sums of classical trajectories and corrections,
\begin{align}\label{sums}
\langle x\rangle=x(t_-)+\langle x_S\rangle,\quad \langle y\rangle=y(t_-)+\langle y_S\rangle.
\end{align}
In the nonrelativistic limit, when $c\to\infty$, the corrections vanish and we obtain the classical trajectories, as is the case of the Schr\"odinger equation.
The corrections (\ref{corr}) to the classical helical orbits are minuscule, of the order of $10^{-14}$m or even less. Therefore, the distortion of classical orbits is totally negligible. The centers of the electron beams described by our solutions of the Dirac equation for all practical purposes follow classical trajectories depicted in Fig.~2.

\section{Conclusions}

The results presented in this work serve a dual purpose. On one hand they add new analytic solutions of the Schr\"odinger and Dirac equations that have a clear physical relevance. On the other hand they may be viewed as explicit examples of the validity of the Ehrenfest theorem \cite{pe,bb4} in relativistic quantum mechanics. In this context we found significant differences between the non-relativistic and relativistic theories. Despite their similarity, these two cases are substantially different. In the non-relativistic case we were able to construct electron wave packets localized in all three dimensions moving along classical trajectories, in accordance with the Ehrenfest theorem. In the relativistic case the wave packets were localized only in the transverse direction. Therefore, they described a beam of electrons extending over all values of $z$, like other known electron beams (Bessel beams, Laguerre-Gauss beams or exponential beams \cite{bb1}). The difference between the predictions of the non-relativistic and relativistic theories, appearing in the analysis of the classical-quantum correspondence, is not surprising. It shows up already at the kinematical level as an energy-dependent lower bound in the Heisenberg uncertainy relations \cite{bb5}. At the dynamical level, the difference is even more pronounced. The Hamiltonian in the Schr\"odinger equation is a direct counterpart of the classical Hamiltonian but the Hamiltonian in the Dirac equation -- a $4\times 4$ matrix -- does not resemble its classical counterpart (\ref{ham2}). It came as a pleasant surprise to find that, despite all the differences, it was possible to find wave functions obeying the Dirac equation which describe beams of electrons with classical helical shapes.

\appendix*
\section{}

The first factor of the unitary operator $\hat{U}$ is a c-number phase factor; it has no effect on the Hamiltonian operator but it affects the time derivative. The second factor leaves the position operators unchanged but shifts the momentum operator by the classical solution and the third factor has the analogous effect on the position operator,
\begin{align}\label{u3}
\exp\left[-\frac{i}{\hbar}p_i(t)\hat{x}^i\right]\hat{p_i}
\exp\left[\frac{i}{\hbar}p_i(t)\hat{x}^i\right]=\hat{p}_i+p_i(t),\\
\exp\left[\frac{i}{\hbar}\hat{p}_ix^i(t)\right]\hat{x^i}
\exp\left[-\frac{i}{\hbar}\hat{p}_ix^i(t)\right]=\hat{x}^i-x^i(t).
\end{align}
Therefore, the action of the operator $\hat{U}$ on the Hamiltonian operator gives,
\begin{widetext}
\begin{align}\label{uh}
\hat{U}^\dagger\hat{H}\hat{U}=\frac{1}{2}(\hat{p}_i-p_i(t))A^{ij}
(\hat{p}_i-p_i(t))+\frac{1}{2}(\hat{x}^i+x^i(t))B_{ij}(\hat{x}^j+x^j(t))
+(\hat{p}_i-p_i(t))C^i_{\;j}(\hat{x}^j+x^j(t)).
\end{align}
The transformation of the time derivative gives,
\begin{align}\label{udt}
\hat{U}^\dagger i\hbar\partial_t\hat{U}
=i\hbar\partial_t+\frac{1}{2}p_i(t)\frac{dx^i(t)}{dt}+\frac{dp_i(t)}{dt}x^i(t)
-\frac{dp_i(t)}{dt}\hat{x}^i -\frac{dx^i(t)}{dt}\hat{p}^i.
\end{align}
\end{widetext}
In the last step we use the equations of motion (\ref{eqm}). One may check now that all terms quadratic and linear in $x^i(t)$ and $p_i(t)$ cancel out in the difference between (\ref{uh}) and (\ref{udt}). Therefore, the equality,
\begin{align}\label{eq}
\hat{U}^\dagger(\hat{H}-i\hbar\partial_t)\hat{U}=\hat{H}-i\hbar\partial_t.
\end{align}
is valid.

\end{document}